\documentclass[conference]{IEEEtran}
\usepackage[obeyspaces]{url}
\usepackage{latexsym}
\usepackage{ngerman}
\renewcommand{\figurename}{Figure}
\renewcommand{\tablename}{Table}
\renewcommand{\abstractname}{Abstract}
\usepackage{graphicx}
\usepackage{caption}
\usepackage{array}
\usepackage{multirow}
\usepackage{hhline}		
\usepackage{subfigure}
\usepackage{mathptmx}
\usepackage{amsfonts}
\usepackage{amssymb}
\usepackage{amsbsy}
\usepackage{adjustbox}
\usepackage{todonotes}
\usepackage{placeins}
\ifCLASSINFOpdf
\else
\fi
\usepackage{url}

\begin{document}
\title{Dragoon: Advanced Modelling of IP Geolocation\\ by use of Latency Measurements\vspace{-0.2cm}}
\author{
\IEEEauthorblockN{Peter Hillmann, Lars Stiemert, Gabi Dreo Rodosek, and Oliver Rose}
\IEEEauthorblockA{Research Center CODE (Cyber Defence)\\
	Universit\"at der Bundeswehr M\"unchen\\
Neubiberg, 85577, GERMANY\\
Email: \{peter.hillmann, lars.stiemert, gabi.dreo, oliver.rose\}@unibw.de
}
}

\maketitle

\begin{abstract}
IP Geolocation is a key enabler for many areas of application like determination of an attack origin, targeted advertisement, and Content Delivery Networks.
Although IP Geolocation is an ongoing field of research for over one decade, it is still a challenging task, whereas good results are only achieved by the use of active latency measurements. Nevertheless, an increased accuracy is needed to improve service quality.
This paper presents an novel approach to find optimized Landmark positions which are used for active probing.
Since a reasonable Landmark selection is important for a highly accurate localization service, the goal is to find Landmarks close to the target with respect to the infrastructure and hop count.
Furthermore, we introduce a new approach of an adaptable and more accurate mathematical modelling of an improved geographical location estimation process.
Current techniques provide less information about solving the Landmark problem as well as are using imprecise models.
We demonstrate the usability of our approach in a real-world environment and analyse Geolocation for the first time in Europe.
The combination of an optimized Landmark selection and advanced modulation results in an improved accuracy of IP Geolocation.
\end{abstract}

\begin{IEEEkeywords}IP Geolocation, Latency Measurement, Lateration \end{IEEEkeywords}

\IEEEpeerreviewmaketitle

\vspace{-0.3cm}
\section{Introduction}\label{sec:introduction}
\textit{Geolocation} describes the process of allocating a physical location, e.g. defined by country, city, longitude, and latitude, to a logical address.
Determining the real-world location of a network entity is called IP Geolocation by using the Internet Protocol (IP) \cite{Padmanabhan2001}.
More and more applications are taking into account from where users are accessing a service.
An important use case are Content Delivery Networks (CDN). 
In this context location information are supporting optimized load balancing between mirror servers and providing improved traffic management, e.g. for downloads \cite{Bindal2006}. 
Another major field of application is location based advertising. 
Customers are automatically redirected to the appropriate language or receive, e.g. advertisement from shops in their surroundings.
Location-aware services offer novel functionalities to their customers and provide adjusted content.
Since criminal actions by use of computers have been emerged for years, whereby quantity as well as the quality of attacks are increasing steadily \cite{koch2012attack}. Determining the geographical location of the attacker is a major concern of law enforcement agencies.
The knowledge of the origination of an attack is mandatory to be able to trace back an attacker and to determine which legal authority is in charge.
Considering law enforcement, location information can support possible pre-forensic and pro-active strategies by classifying traffic, for instance according to the source or destination country.
An additional use case is the examination of the network infrastructure, e.g. to detect routing anomalies or rerouting 
as well as which routing policies have been applied. 
Further areas of application are prevention of credit card fraud and content restriction due to political meaning or licensing \cite{EAR2010}.
%
The necessity for a highly accurate and reliable Geolocation service has been identified as an important goal for the future Internet. 

\begin{figure}[bhpt]
	\centering
	\captionsetup{justification=centering}
	\includegraphics[width=0.47 \textwidth]{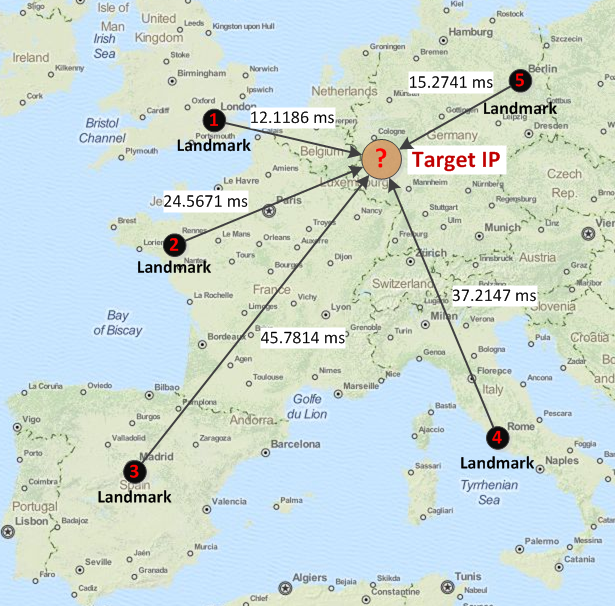}
	\caption{Example of measurement based IP localization.}
	\label{fig:GeneralMeasurement}
\end{figure}

In general Geolocation can be conducted by applying IP mapping and measurement based strategies. 
Approaches which are relying on IP mapping, determine locations by retrieving information about an IP address from databases and are considered as passive in terms of communication with the target. As illustrated in Fig. \ref{fig:GeneralMeasurement} measurement based techniques are relying on actively probing a particular host and infer the geographical location by measuring latencies.
For this purpose most of these procedures are utilizing reference hosts with well-known location information, called \textit{Landmarks} or \textit{Vantage Points}. Since these approaches are using the moderate correlation between network delay and geographic distance, the accuracy is mainly influenced by the selection of Landmarks and the mathematical modelling \cite{Ziviani2005503}.

The paper introduce two novelties. First, we present Dragoon an improvement in terms of positioning and selection of Landmarks for Geolocation strategies based on latency measurements.  
Second, we describe a new approach of an advanced and more accurate mathematical modelling of the geographical location estimation process.
In addition, misleading statements of other publications like the propagation speed of signals are clarified.
Therefor, the correlation between network delay, latency measurement, network topology, and geographical distance is analysed for the first time focusing Europe.


\section{Scenario and Requirements}\label{sec:scenario}
The need for an accurate and reliable Geolocation service is illustrated by using the following real-world scenario. A sophisticated attack on a company network is detected by a behaviour based Intrusion Detection System (IDS). The management of the company requires a clarification of the case. As part of a pre-forensic strategy, the physical location of the attack source should be determined. Therefor, a Geolocation service is necessary, which uses the determined IP address of the possible attacker and correlates delay pattern with the estimated distance. 
The extracted information are used to prove the evidence and to start further actions. Beside this, all data is recorded and stored with signatures to cope the forensic needs of law enforcement. 

\begin{figure}[hbtp]
{
\centering
\captionsetup{justification=centering}
\includegraphics[width=0.47 \textwidth]{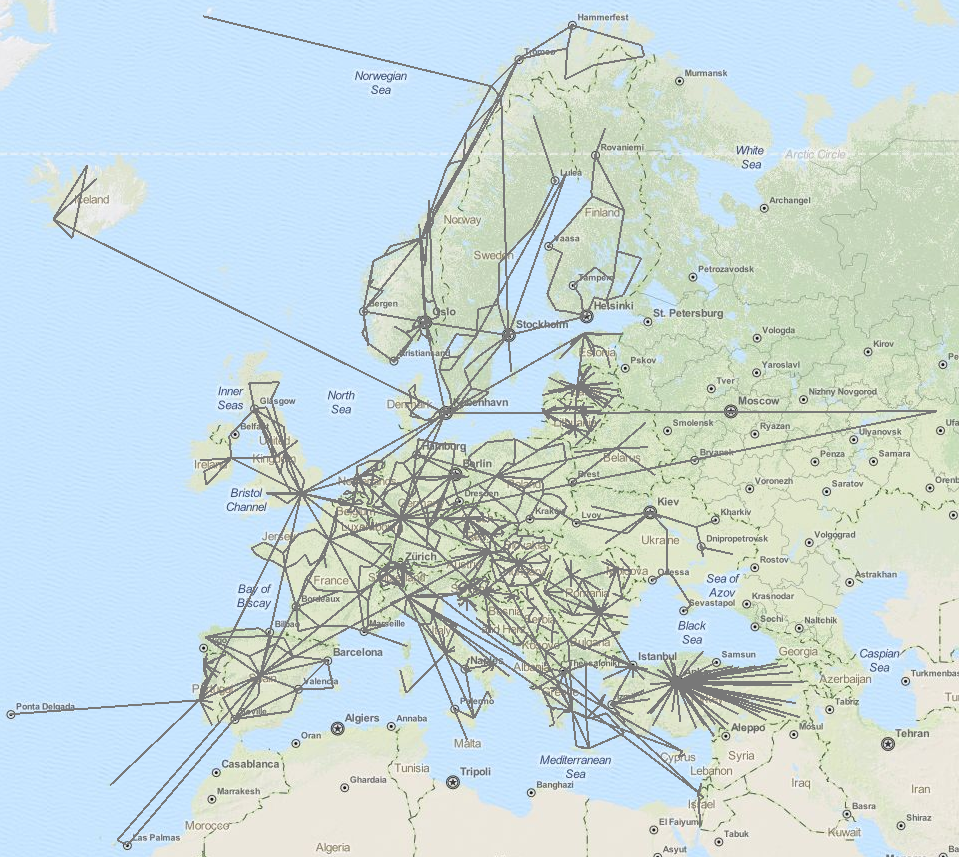}
\caption{Example Internet backbone network topology in Europe \cite{Knight2011}.}
\label{fig:backboneTopology}
}
\vspace*{-0.6cm}
\end{figure}

Since the location of the aggressor is not known beforehand, a well distributed network of multiple Landmarks is required.
The chosen Vantage Points have to be close to the measured target with respect to the infrastructure and the hop count. As closer such a reference host is to the target, the impact of interfering influences during the active latency measurement is reduced. Because of the high-performance connection, the ideal case are Landmarks placed within the backbone network topology. 
The accuracy of latency measurements and the optimal selection of Landmarks is directly related to the precise detection of a nearby target. 
The amount of Landmarks needed is mainly influenced by the network load due to applied multiple measurements and cost-efficiency in terms of maintaining a probing infrastructure.
An example of such an optimization problem is shown in Fig. \ref{fig:backboneTopology}. It presents the backbone network in Europe, where we need to find a predefined number k center nodes with minimized maximum distance to the surrounding network topology. This optimization problem is NP-hard 
and is based on the classical k-center problem for clustering. Therefor, we introduce a novel strategy to find a predefined number of optimized positions in the top level network topology with known geographical coordinates. 
Based on optimized Landmark selection and positioning as well as the detailed model, latency measurements by use of Round-Trip-Times (RTT) are conducted.
%

\section{Related Work}\label{sec:sota}
\subsection{IP Geolocation}
According to Endo et al. \cite{Endo2010} approaches for IP Geolocation can be classified in either IP mapping based - including semantic - or measurement based strategies.

%
IP mapping based along with semantic strategies determine locations by basically retrieving information about a given IP address based on queries against databases. Examples of such approaches are the Domain Names System (DNS) LOC resource record, 
querying a certain Regional Internet Registry (RIR), 
by conducting WHOIS-lookups, and Geolocation databases offered by geoservices like MaxMind \cite{Maxmind}.
%
%
%
Specific implementations
are NetGeo \cite{Moore2000} and Structon \cite{Guo2009} as well as GeoCluster and GeoTrack of the IP2Geo suite \cite{Padmanabhan2001}. 
%
%
Current IP mapping based approaches - particularly Geolocation databases - have been proven to be accurate up to 98\% on a country level \cite{Poese2011}. 
Although location information obtained by querying the five RIRs has been considered as questionable \cite{Endo2010}. Current findings have shown the need for a comprehensive evaluation. 
If existent, the DNS LOC resource record provides accurate longitude and latitude data. Due to missing acceptance it has not been widely adopted. In addition, the accuracy is depending on the network operator and company restrictions \cite{Gueye2006}. 
When it comes to highly accurate results, IP mapping based approaches can be considered as coarse-grained, providing only rough and incomplete location information. 

In this work, we focus on the more dynamic and actual measurement based techniques.
%
These are relying on an active interaction with the target system. This can be achieved by sending various Internet Control Message Protocol (ICMP) echo requests and recording the amount of time it takes for the response to arrive. Basically this is conducted by using ping or traceroute. The similarity of all those strategies is, that they are based on the assumption of an existent correlation between network latency and geographical distance.
%

\textit{Shortest Ping} \cite{Katz-Bassett2006} is a simple delay-based technique. Each target is mapped to the Landmark that is closest to it in terms of the measured RTT.
\textit{GeoPing} as part of IP2Geo \cite{Padmanabhan2001} uses network delay measurements from geographically distributed locations to set up latency vectors. 
The corresponding location of the latency vector which is most similar to the target is inferred as geographic location. 

In addition it is possible to further divide measurement based techniques in constrained- and topology-based as well as hybrid approaches. Since hybrid approaches might be capable of integrating also IP mapping based strategies, they are still basically relying on active probing and thus can be considered as measurements based.
%
\textit{Constraint-Based Geolocation (CBG)} \cite{Gueye2006} infers the geographic location of Internet hosts by using multilateration with distance constraints. Hence establishing a continuous space of answers instead of a discrete one.
\textit{Topology-Based Geolocation (TBG)} \cite{Katz-Bassett2006} introduces topology measurements to simultaneously geolocate intermediate routers. Nevertheless TBG is only an enhanced version of the original CBG.

To overcome possible shortcomings and to verify the results, more and more hybrid solutions are developed \cite{Endo2010}.
%
\textit{Octant} \cite{Wong2007} is a framework for IP Geolocation and the current ``State-of-the-Art'' in terms of active measurements based approaches \cite{Eriksson2011}. The approximate host location is inferred by using geometric regions and thus a continuous solution space. Therefor, \textit{B\'ezier Curves} as well as positive and negative constraints are applied to limit the possible geographic area. By using B\'ezier Curves large and complex areas can be represented in precise way. 
Octant has a modular design, hence being capable of using additional constraints like demographic data to improve the accuracy. 
%
Other examples for hybrid or measurement based approaches are HawkEyes \cite{Dahnert2011}, Spotter \cite{Laki2011} and Posit \cite{Eriksson2011}.

A survey of most important work in the field of Geolocation is provided by the Center for Applied Internet Data Analysis (CAIDA) \cite{CAIDA2015}. The most accurate results are obtained by measurement based approaches, with up to 600 meters close to the target \cite{Wang2011}. But these results are carried out by conducting the proposed techniques in either research environments with homogeneous infrastructures or by applying assumptions like a company hosts their own webserver in-house. In times of Cloud-based and outsourcing strategies, a company may not host its service locally. They use CDNs to distribute content or use shared hosting techniques. In this case, there is no direct one-to-one mapping between an IP address, the service providing company, and the corresponding location.

\subsection{Landmark Problem}\label{landmarkProblem}
Almost all current measurement approaches, in particular the obtained accuracy, are highly depending on Landmarks. The dilemma of using as much as necessary but as few as possible and well distributed Landmarks to reach a highly accurate location estimation, can be considered as \textit{Landmark Problem} \cite{Eriksson2011}. 

%

All notably measurement based approaches use Landmarks for their analyses, but do not provide comprehensive information about optimal selection, positioning or give no attention on this problem at all. In addition, they are using euclidean distances. The work from Ziviani et. al \cite{Ziviani2005503} provides an algorithm for placement of Landmarks. Their proposed linear programming (LP) model is used as a reference for the Landmark location selection. Nevertheless, the presented approach is not suitable for realistic and large scale scenarios. 
Thus, our focus is to determine a predefined amount of Landmarks in the given infrastructure to improve the measurement based IP localization.

\section{Concept}\label{sec:concept}
Our proposed Geolocation service uses a predefined amount of known Landmarks to actively probe the target IP address or network device one hop before. The RTT and hop count from all Landmarks to the target are measured for geographical distance correlation and further calculations. Through multilateration by using these results, the geographical location is inferred. 
The underlying problem is the selection and placement of Landmarks out of a given set of possible locations for probing. The Vantage Points have to be identified automatically in the topology in respect to minimize the maximum distance to the surrounding network topology. With an improved placement of Landmarks, these are closer to the target. As nearer they are, the distance is lower and the variance of measurement results is reduced.


\subsection{Dragoon: Finding Landmarks}\label{sec:landmarks}\label{sec:algorithm}
Based on the k-Means clustering strategies \cite{MacQueen67}, we introduce a new algorithm Dragoon (Diversification Rectifies Advanced Greedy Overdetermined Optimization N-Dimensions) to find optimized locations for Landmarks, which represent central nodes in a given network topology. Moreover, the established algorithms are very sensitive to an initial solution. The first placed Landmark usually covers a high amount of nodes, which shows the serious influence of the first placement decision. Nevertheless, an even distribution of Landmarks would be desirable to measure from different directions using several network paths. The mentioned approaches try to find optimized locations only with respect to these clustered groups. The influence to other groups and the overall system is lost, leading to suboptimal solutions. As the target location is not known beforehand, a distributed network of Landmarks is highly recommended.
%
With the knowledge of these problems and weaknesses of current solutions, we design our own algorithm. 

After an initialization, the network nodes are assigned to the closest Landmark location. In an iterative optimization these locations are improved.
To reduce the sensitivity to the first placed Landmark of the k-Means strategies, we design a novel initialization process which avoids the influence of random decisions in order to get stable solutions. In the preliminary stage, an orientation mark is placed at the optimal position to the distances of the given network topology. Afterwards, the specified amount of Landmarks is placed using the 2-Approx strategy \cite{Gonzalez85}. 2-Approx calculates for every network node the distance to all placed Landmarks. It chooses the node with the largest distance to their closest Landmark as the new location to place the next Landmark. Thereby, we obtain a specific solution of the 2-Approx placement strategy, which guarantees the 2-approximable quality of the result. 
This ensures that the maximum value of a distance from a network node to its closest Landmark is not larger than twice the maximum considering the optimal placement location of all Landmarks.

\begin{figure}[hbtp]
\vspace*{-0.1cm}
{
\centering
	\captionsetup{justification=centering}
\includegraphics[width=0.47 \textwidth]{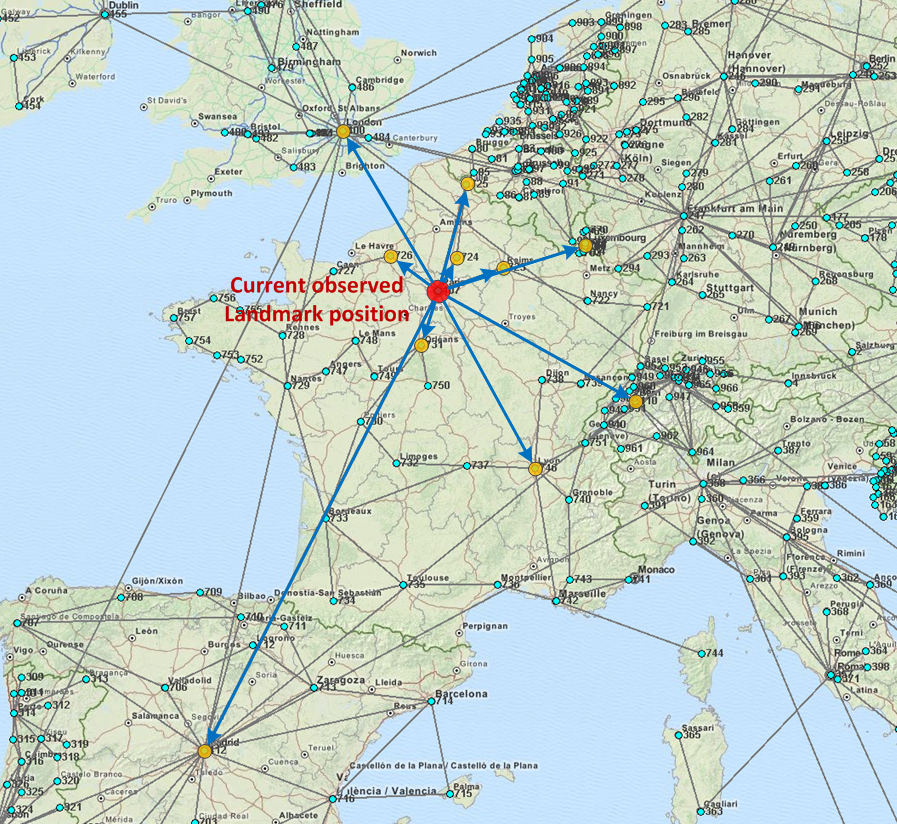}
\caption{Improvement of the current Landmark position (red) with the possibilities (yellow) tried by Dragoon and considered other network nodes (blue).}
\label{fig:DragoonImprovement}
}
\vspace*{-0.3cm}
\end{figure}

After the initialization, the algorithm starts with the iterative refinement to find the list of final locations for Landmarks. The following description explains the specific approach for the network infrastructural placement, which is adaptable to other constraints. The algorithm checks all possible locations around the observed Vantage Point and tests all connected nodes with a direct edge to the current position, see Fig. \ref{fig:DragoonImprovement}. If the new location improves the overall situation, the algorithm accepts it and replaces the current observed Landmark with the new position. This is done with respect to the specified optimization criterion. In our case, it is the maximum distance counted by hops. If this value is unchanged, the algorithm will use an additional criterion. We use an average or mean criterion to choose between two solutions and to identify an improvement. In each iteration step, the network nodes are (re)assigned to their closest Landmark. 
All actual Landmark locations are used in each evaluation step - except of the observed one. In each iteration, each Landmark is allowed to shift its position only once. This leads to a stepwise improvement and avoids a too fast stagnation in a local optimum. Due to the global view, the order of the selected Vantage Points for the stepwise optimization has almost no influence on the final result. The Landmarks are selected in the same order as they are added to the network topology. This iterative optimization is repeated until all Landmark positions do not change any more. Due to the described initialization, only a few iterations are necessary until the algorithm terminates. The algorithm accepts only improved positions in every step. Therefore, the 2-approximable condition holds and it will always terminate.

The algorithm can also be adapted to extend an existing Landmark infrastructure or other use cases. An additional area of application is the placement of mirror servers for CDNs in a given network topology.
%
 In comparison to other Landmarks selections procedures, we achieve a better distribution of measurement points with a shorter estimated distance to the target.
 
%
%
%

\subsection{Conversion of RTT and hop count to distance}\label{conversionRTT}
Prior work has shown that packets travel in fiber optic cables between $\frac{ 2 }{ 3 }$ \cite{Percacci2002}  and $\frac{ 4 }{ 9 }$ \cite{Katz-Bassett2006} of the speed of light in vacuum (c). Pure physical calculation results in about 200 000 km/s for glass with a refractive index of 1.5. Theoretical information transmission in copper is about $\frac{ 3 }{ 4 }$ c $\approx{}$ 225 000 km/s, 
higher than in glass. Based on this knowledge, the latency has to be measured in microseconds to limit the basic time measurement error below 500 meter. In comparison, general measurements in milliseconds lead to error variations within about 800 km. Nevertheless, due to transmission delay, queuing delay and further processing influences these theoretical speeds are not applicable in real-world scenarios. It just give us an intention, how precise the time has to be measured.

As we know from the CAIDA data \cite{CAIDA20152} and Section \ref{sec:sota}, the correlation between latency and real distance follows approximately a logarithmic curve. Eq. \ref{logarithmicCurve} presents the formula of such a curve fully parametrized, whereas $latency$ is $\frac{RTT}{ 2 }$ subtracted by the $average$ $delay$ of the $detected$ $hops$, which is about $0.1$ $ms$ per hop.  
The reason for the logarithmic correlation is the relatively large transmission delay through the processing units compared to the signal propagation speed in the conductor. This influence is particularly strong in the so called ``Last Mile'', the connection to the end user. 
In comparison, most current research work abstract this correlation as a linear function. Such modelling do not take the different Tier network levels into account. In this case it can be considered to be imprecise.
\begin{equation}
distance = p * log_{ e }\left( q * latency + n \right) + m
\label{logarithmicCurve}
\vspace*{-0.1cm}
\end{equation}

The parameter p, q, n and m for such a curve are not known. Because of their unique location in the network topology, they have to be calculated for every Landmark individually. For the curve reconstruction and evaluation, we estimate the function based on multiple inter Landmark measurements and curve fitting with a minimized sum of squared-error. As we select Landmarks in the backbone topology with high-performance connection, the resulting curve will be too optimistic. Therefor, we counter this by considering the hop count and applying an adaptable training factor, called $LC$. This factor allows us a general shifting of the logarithmic curve to the top or bottom and with it the estimated distances.  

The distance calculation in our model is based on the WGS84 reference ellipsoid as well as orthodromic distances, also known as great-circle. This is also used by the Global Positioning System (GPS). Thereby, we achieve a much higher accuracy than modelling the earth as a ball without mountains and valleys, rotation flattening effect and applied euclidean distances. 
An orthodromic distance corresponds to  the shortest distance between two locations on the surface of a sphere, whereas the euclidean space represents the length of a straight line between these locations. Indeed, the path on the surface especially along the network infrastructure is obvious longer than a straight line, which leads to larger distances and wrong estimated target locations. Furthermore, as the entire topology is not known, the influence can still be estimated with the knowledge of the typical installation of cables along roads through practical construction. Therefor, an evaluation with distances calculated by Google Maps are analyzed and compared with orthodromic modelling.



\subsection{Lateration}
A geographical location of a target can be estimated using the measurement method of lateration. It uses two known geographical Landmark locations and the estimated distance from each to the target. To increase the precision, multiple Landmarks are probing the targets IP address. As we dealing with distances by the calculation of lateration, the measured RTT and the hop count are converted to a distance using the determined logarithmic function and the principle described in Section \ref{conversionRTT}. Thus, the location of each Landmark and the distance from there to the target are known. The calculated distance represents the radius $r_{t}$ of a circle or ellipse with the location of Landmarks in the center ($x_{t}$,$y_{t}$), see Eq. \ref{ellipseFunction}:
\begin{equation}
\left( x - x_{ t } \right)^{ 2 } + \left( y - y_{ t } \right)^{ 2 } = r_{ t }
\label{ellipseFunction}
\vspace*{-0.2cm}
\end{equation}

For the lateration, we calculate the intersection of two circles. This is done without using angles and is mandatory for a precise modelling, due to the shape of the Earth. The following quadratic Eq. \ref{formula1} shows the general solution for the intersection of two circles, dissolved after the y value, where y correspond to Eq. \ref{formulaY}. For clear presentation, the partial Eq. \ref{formulaH1} to \ref{formulaH4} are separately listed.
\begin{equation}\label{formula1}
\mathbf{y'^{ 2 }} - \frac{ (a + b ) * d }{ c^{ 2 } + d^{ 2 }} *\mathbf{y'} - r_{ 1 }^{ 2 } - \left( \frac{ a+b }{  2*c } \right)^{ 2 } = 0
\end{equation}
\begin{equation}\label{formulaY}
y = y' + y_{1}
\end{equation}
\begin{equation} \label{formulaH1}
a = x_{ 1 }^{ 2 } + y_{ 1 }^{ 2 } - r_{ 1 }^{ 2 } - x_{ 2 }^{ 2 } - y_{ 2 }^{ 2 } + r_{ 2 }^{ 2 }
\end{equation}
\begin{equation} \label{formulaH2}
b = -2 \left( x_{ 1 } - x_{ 2 }\right) *x_{ 1 } - 2\left( y_{ 1 }  - y_{ 2 }\right) * y_{ 1 }
\end{equation}
\begin{equation} \label{formulaH3}
c=\left( x_{ 1 } - x_{ 2 } \right)
\end{equation}
\begin{equation} \label{formulaH4}
d=\left( y_{ 1 } - y_{ 2 } \right)
\vspace*{-0.1cm}
\end{equation}

Assuming that up to two solutions for $y$ can arise, due to the circle equation each one in turn can have up to two solutions for $x$. The subsequent check in both intersecting circles reduces the four solutions to the expected, maximal two intersection points.

\begin{figure}[hbt]
\begin{minipage} [hbt]{4.192cm}
\vspace*{2mm}
\begin{center}
		\captionsetup{justification=centering}
\includegraphics[width=1.19 \textwidth]{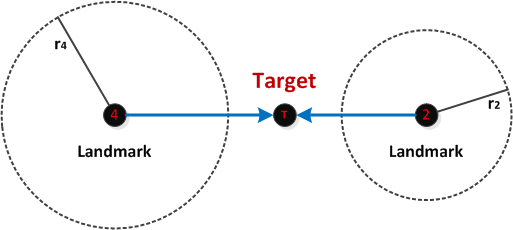}\\
\caption{Zero intersections.\newline Non-overlapping circles.}
\label{fig:circleIntersection0}
\end{center}
\end{minipage}
\hspace*{8mm}
\begin{minipage} [hbt]{4.192cm}
\begin{center}
		\captionsetup{justification=centering}
\includegraphics[width=0.59\textwidth]{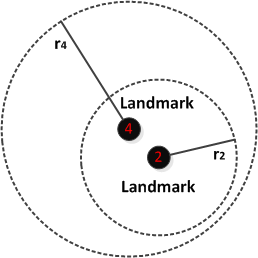}\\
\caption{Zero intersections.\newline Overlapping circles.}
\label{fig:circleIntersection1}
\end{center}
\end{minipage}
\end{figure}

\begin{figure}[hbt]
\begin{minipage} [hbt]{4.192cm}
\vspace*{5mm}
\begin{center}
		\captionsetup{justification=centering}
\includegraphics[width=1.03 \textwidth]{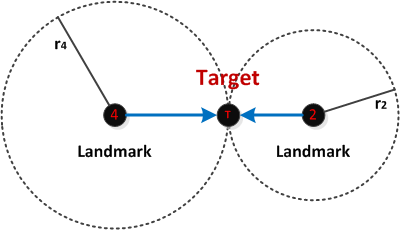}\\
\caption{One intersection point.}
\label{fig:circleIntersection2}
\end{center}
\end{minipage}
\hfill
\begin{minipage} [hbt]{4.192cm}
\begin{center}
		\captionsetup{justification=centering}
\includegraphics[width=1.0 \textwidth]{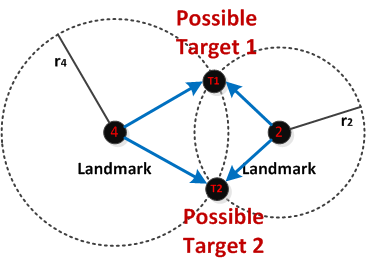}\\
\caption{Two intersection points.}
\label{fig:circleIntersection3}
\end{center}
\end{minipage}
\end{figure}

Two circles can have zero, one or two intersections. In the case of zero intersection points and non-overlapping circles, the target location can be estimated in the middle of the space between two Landmarks. If the circles are completely overlapping it is likely, that the target is in the range of the circle with the smaller radius. Otherwise, results from zero intersection point can be disregarded in further calculation. Alternatively, we reduce the radius of the larger circle until we get an intersection point. As we deal with probabilities, the estimated location is likely to be at this point. In the case of just one intersection point we assume this point as the location of the target. The only inconclusive case is about two intersection points. Here we need further information calculated with support of other Landmarks to decide between the two possible solutions to get the right one. 
The different cases are visualized in the Fig. \ref{fig:circleIntersection0} to \ref{fig:circleIntersection3}.

Since the correlation depends on the region and their connectivity as well as on the used network components and materials, determining reasonable Landmark positions is an important challenge for a highly accurate IP localization service.
\vspace*{-0.1cm}
\subsection{Target location estimation}
\begin{figure}[hbt]
\begin{minipage} [hbt]{2.292cm}
\begin{center}
		\captionsetup{justification=centering}
\includegraphics[width=0.98 \textwidth]{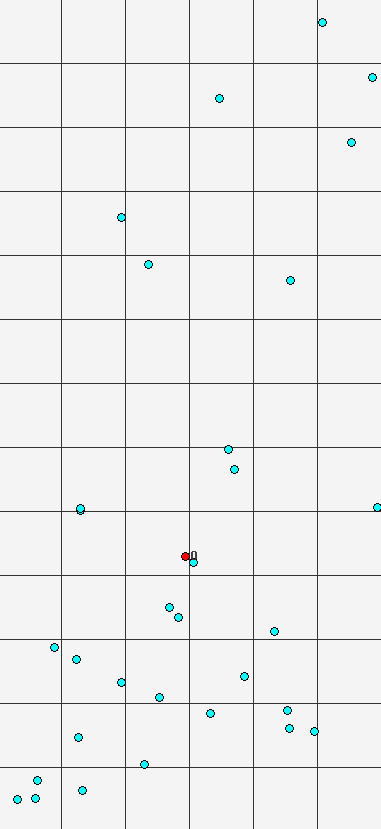}\\
\caption{Intersection point cloud.}
\label{fig:cloud}
\end{center}
\end{minipage}
\hfill
\begin{minipage} [hbt]{6.092cm}
\begin{center}
		\captionsetup{justification=centering}
\includegraphics[width=1.02 \textwidth]{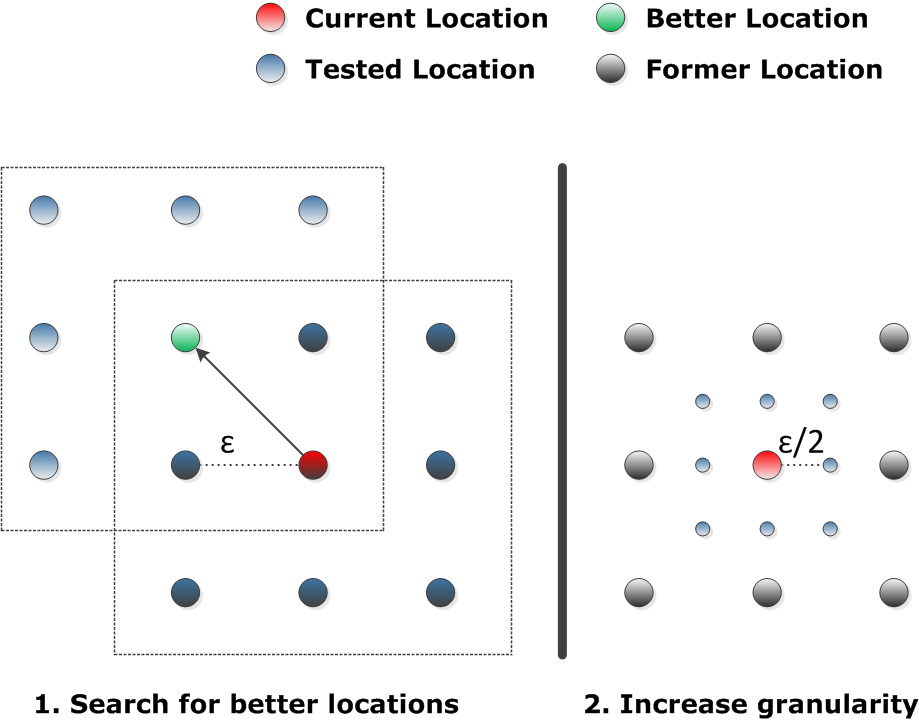}\\
\caption{Iterative optimization stage of the algorithm Dragoon by free placement constraint.}
\label{fig:DragoonImprovement2}
\end{center}
\end{minipage}
\vspace*{-0.4cm}
\end{figure}

As result of the conducted multilateration, we get a cloud of multiple locations, where the target location is estimated as shown in Fig. \ref{fig:cloud}. With our Dragoon algorithm, adapted to a constrained free center placement, we calculate the center location of all points according to the optimization criterion \textit{minimized average distance}. For the free placement constraint, our algorithm tests all points on a grid with a defined distance ($\epsilon$). If one of the tested locations results in a better performance for the overall scenario, it will be accepted. This location is used for the next iteration step. If no location leads to an improvement, we successively decrease the granularity of the grid \mbox{($\epsilon_{new}$ := $\frac{\epsilon_{old}}{2}$)}. This process is repeated until the grid distance $\epsilon$ is smaller than the maximal accepted deviation. 
The processing steps of the iterative optimization are shown in Fig. \ref{fig:DragoonImprovement2}. The left side illustrates the movement to an improved spot. The right side shows the increased granularity of the grid by bisection. For an improved target location estimation, we filter single points, which are too far away from the other points. Therefor, we iteratively repeat the placement of the center node and filter the points with largest distances from the optimized center location.

\subsection{Measurement}
From a measurement point of view, the end-to-end delay over a fixed path can be split into two components: A deterministic and a stochastic delay \cite{Bovy2002}. The deterministic delay is composed by the minimum processing time at each router, the transmission delay, queuing delay, and the propagation delay. This deterministic delay is fixed for any given path and is taken into account in our concept described in Section \ref{conversionRTT}. The stochastic delay composes the queuing delay at the intermediate routers and the variable processing time as well as buffering at each router that exceeds the minimum processing time. To counter this stochastic delay, several measurements are necessary to get a value close to the theoretical minimal RTT. 
Considering the multiple results, we are only interested in the minimal RTT to get the correct distance and to avoid misleading measurement values caused by circuitous routing. 
Apart from the delay measurements, the hop count on the path has to be determined by tracing the target. 
This value is used by time measurement and in the calculation in Section \ref{conversionRTT}.

\section{Evaluation}\label{sec:evaluation}
In order to verify the improvements of our introduced concept and algorithm, in terms of selection and positioning of Landmarks, we set up an experiment for geolocating IP addresses.
\subsection{Evaluation Environment and Procedure}
To get a first impression how our algorithm is performing we are focusing on Europe using public available information as well as research data on the Tier 1 topology like \cite{Knight2011}. The backbone network in Europe is in comparison to less industrialized parts of the world fully developed, hence providing a reasonable test environment. Since the introduced algorithm calculates the optimal position for Landmarks in respect to a given topology, we have to build a set of distributed reference hosts to which we have access to. This is mandatory for probing and determining the geographic location of our target systems. For this purpose we are using the RIPE Atlas Project \cite{Ripeatlas} providing us with over 8200 well-known nodes for probing, including about 120 so called anchors. Anchors are special nodes with more capacity and are often placed at well connected places in the backbone infrastructure. Furthermore, the exact geographic location of these anchors is provided by the RIPE Atlas Project. 
To avoid confusion the calculated Landmarks are in the following referenced as \textit{Center Nodes}, wheres the reference hosts which we are in the end using for probing are the actual Landmarks. For a first evaluation, we used ten Landmarks to cover entire Europe.

The first step is to calculate optimal positions, determined by latitude and longitude, in respect to the given topology. Afterwards we compare these Center Nodes to our set of reference hosts to find direct matches according to latitude and longitude. If no direct match is possible, we chose the reference host as actual Landmark, which is closest to the position of the calculated Center Node.
The next step is to measure the RTT and hop count between all Landmarks identified in the previous step. To determine the hop count and the RTT we use ``Paris Traceroute'' and ICMP echo request provided by the RIPE Atlas measurement interface. By using the hop count and the measured minimum delay out of ten measurements, a logarithmic curve is calculated in order to represent a correlation between measured latency and geographic distance. The curve reconstruction is calculated parameter pairwise iteratively with the tool R and the curve fitting method $nls$.

\begin{figure*}[htbp]
	\centering
		\captionsetup{justification=centering}
	\includegraphics[width=0.93 \textwidth]{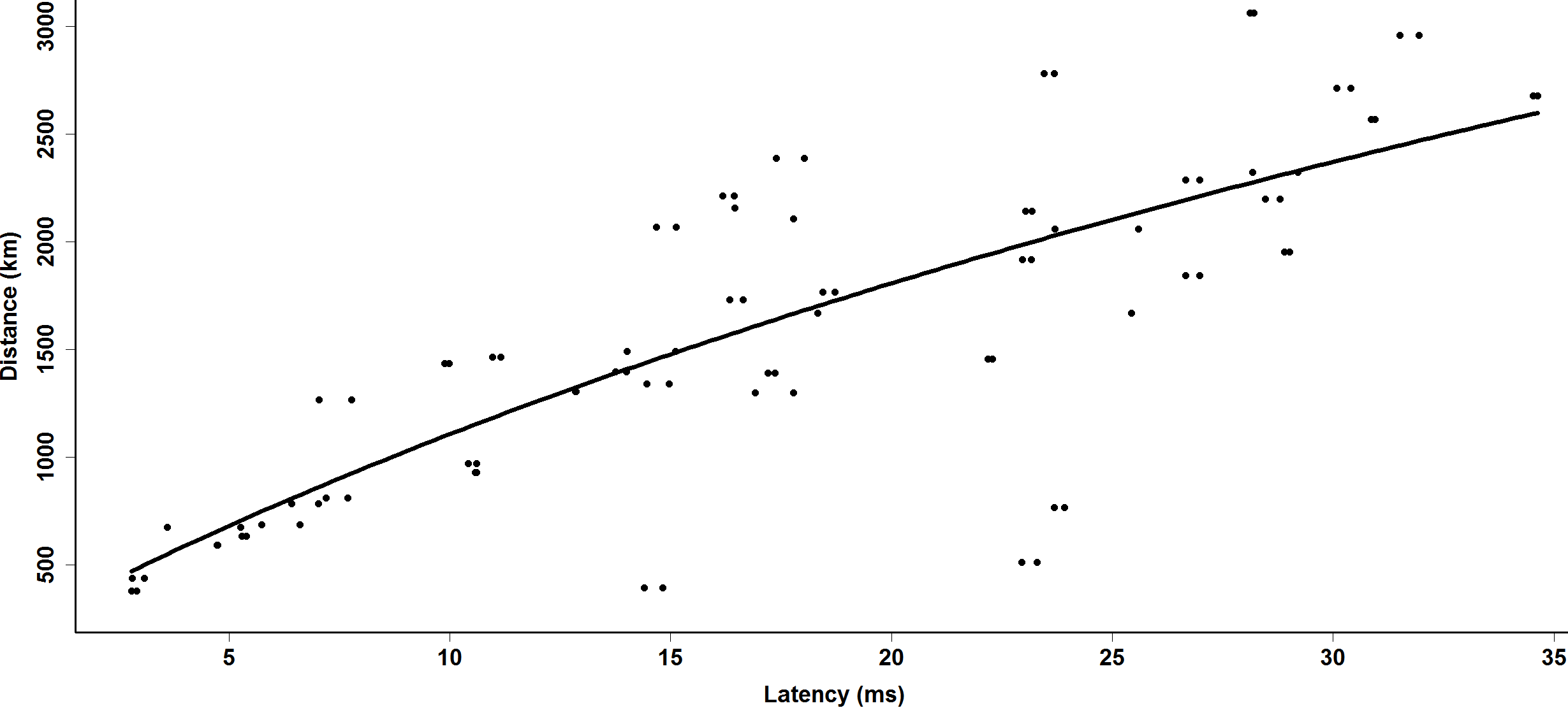}
	\caption{Correlation between Latency and Google Maps distance}
	\label{fig:rtt2distanceGoogle}
\end{figure*}

\begin{figure*}[htbp]
	\centering
		\captionsetup{justification=centering}
	\includegraphics[width=0.93 \textwidth]{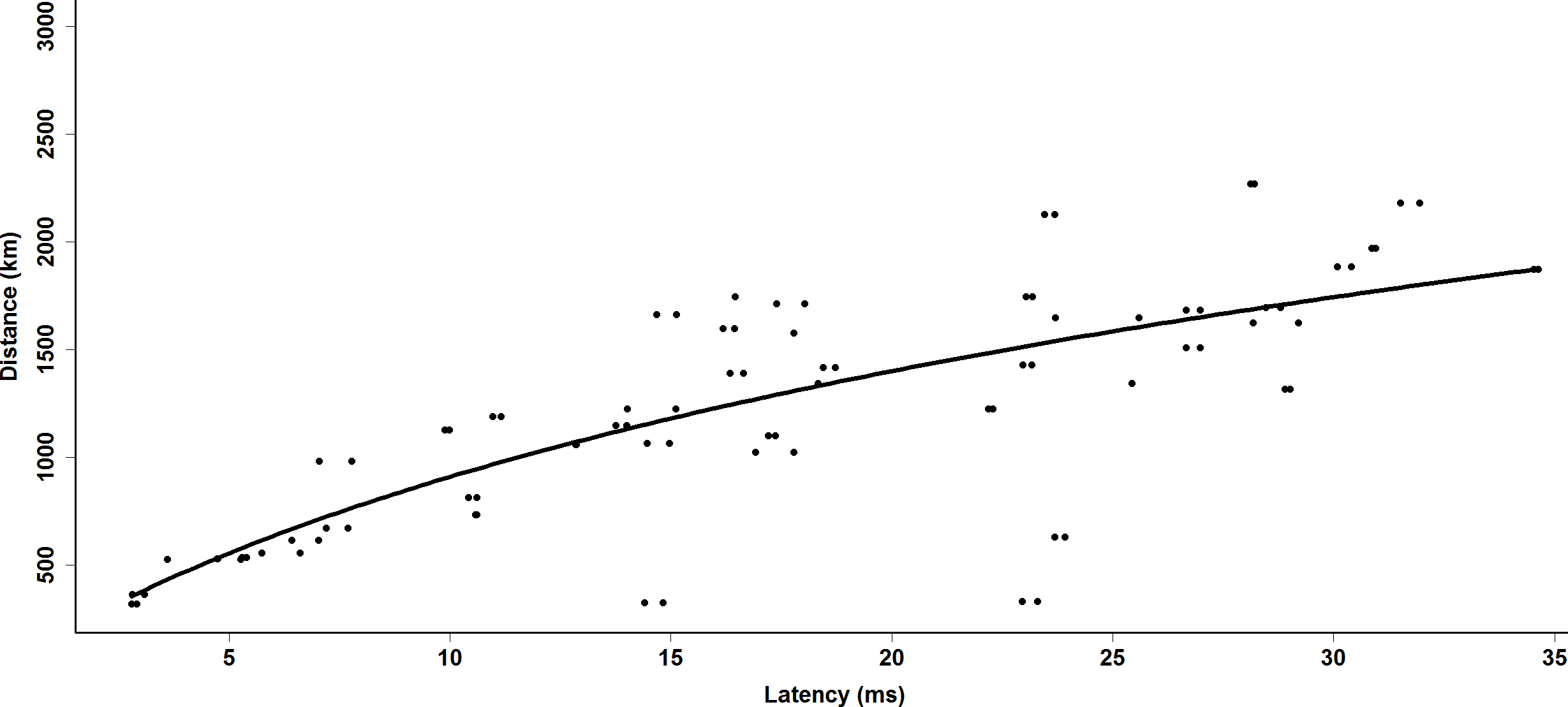}
	\caption{Correlation between Latency and orthodromic distance}
	\label{fig:rtt2distanceOrthodrom}
	\vspace*{-0.3cm}
\end{figure*}

Fig. \ref{fig:rtt2distanceGoogle} and \ref{fig:rtt2distanceOrthodrom} show the difference between Google Maps and orthodromic distances using optimized Landmarks. The function using orthodromic distances is more curved and visualizes the impact of the ``Last Mile'', which is imprecise
covered by the modelling. For this reason and because of improved and more stable location estimation, the further calculations are based on the curve using Google Maps. After probing the target IP address from each Landmark, the curve is used to convert the RTT and hop count to a geographic distance. Using the calculated distance and the knowledge of the longitude as well as the latitude of each probing Landmark, Dragoon is able to infer the actual location of the target.

\subsection{Results and Findings}
The Tab. \ref{tab:AlgoCompare} shows an excerpt from the comparison of estimated target locations obtained by different applied Landmarks, which are identified by Dragoon and 2-Approx.
Since the used scenario is too large for common LP solver, we used the alternative algorithm 2-Approx to the LP in Section \ref{landmarkProblem}. It illustrates the impact of the selected Landmarks to the results of IP Geolocation.

As we know that the logarithmic curve is too optimistic, we tried to find a uniform $LC$ factor to counter it. For the curve based on Landmarks optimized by Dragoon the $LC$ factor is about $0.7$ for distances obtained by Google Maps (GM) and Orthodroms (OD), which means the distance to a given Latency is \mbox{30\%} to large. For the function based on Landmarks identified by the reference algorithms the $LC$ factor is $0.9$.
With an optimized $LC$ factor per target, we are able to achieve more precise results (see Tab. \ref{tab:AlgoCompare2}). This justifies our assumption that the function has a strong influence on the result. Because of the unique location of each Landmark in the network topology, it is further necessary to determine an individual curve for each of them. This will enhance the accuracy additionally. A training data set would improve the curve as well as the approximated $LC$ factor.
Nevertheless, in comparison to the solution presented in \cite{Laki2011} and \cite{Eriksson2011}, 
we achieved better results based on active measurements.
Considering the stochastic delay, the measurements have been conducted between afternoon and early evening. During this time the network load and variance is higher in comparison to other day times. Nevertheless, our applied modelling shows stable results.

\begin{table}[hbtp]
	\centering
		\captionsetup{justification=centering}
	\caption{Comparison of the derivation between the location estimation to the real geographic location using different Landmarks.} 
		\label{tab:AlgoCompare}
	\begin{adjustbox}{max width=\textwidth}
		\begin{tabular}{clll}
			\hline
			\textsc{Target} & \textsc{Dragoon (GM)} & \textsc{Dragoon (OD)} & \textsc{Reference Algorithm}\\
			\hline
			1 & 120 km& 117 km & 350 km \\
			2 & 136 km& 536 km & 1600 km \\
			3 & 221 km& 108 km & 113 km\\
			\hline
		\end{tabular}
	\end{adjustbox}
\end{table}

\begin{table}[hbtp]
	\vspace*{0.2cm}
	\centering
		\captionsetup{justification=centering}
	\caption{Derivation between the location estimation to the real geographic location using adapted $LC$ factors.}\label{tab:results_deri}
		\label{tab:AlgoCompare2}
	\begin{adjustbox}{max width=\textwidth}
		\begin{tabular}{ccc}
			\hline
			\textsc{Target} & \textsc{Dragoon (GM)} & \textsc{\parbox{4.0cm}{\centering Curve distances multiplied by\\ ($LC$ factor)}}\\
			\hline
			1 & 91 km & 0.6 \\
			2 & 85 km & 0.8 \\
			3 & 48 km & 0.5 \\
			\hline
		\end{tabular}
	\end{adjustbox}
\end{table}

\vspace*{-0.2cm}
\subsection{Limitations}
If the underlying infrastructure is unknown a determination of Center Nodes is not possible. Also helix like infrastructures will result in suboptimal calculations, hence more coarse-grained location estimations. Since it has to be assumed that at least the Tier 3 infrastructure provides in comparison to Tier 1 less capacity and connectivity, the logarithmic curve based on the Tier 1 infrastructure has to be considered optimistic. To overcome these shortcomings a more comprehensive knowledge of the Tier 1 to 3 infrastructure, the ``Last Mile'' as well as the transmission medium and the network load of different components is needed. Thus, the amount of Landmarks needed for probing is highly depending on that knowledge. Further factors which may have influence on the location estimation are the used protocol for probing. Traceroutes can be done by different protocols and algorithms, hence causing more or less overhead while different processing steps on the packet path. Caused by its design IPv6 may have impact on the measurements results, too.


\section{Conclusion}\label{sec:conclusion}
In this paper we propose a novel strategy Dragoon to optimize the selection and positioning of Landmarks in a given network infrastructure. Our strategy outperforms existing Geolocation approaches based on active measurements in real-world environments. Considering the selection and positing of Landmarks, our algorithm achieve results close to the global optimum. We show the general usability of time measurements for IP localization with high precision based on the selection and position of Landmarks. Reasonable Landmark positions are important for an accurate IP localization service. The closer a Landmark is to a target, the lower are the interferences during the measurements.

In further research work, we will show that the accuracy can be improved with detailed knowledge about the entire infrastructure between Landmarks and the target node. Combined with packet tracking, the entire network path becomes visible and the correlation between round-trip delay and distances can be determined more exactly. We are currently investigating how Dragoon is performing in moderately connected Internet regions as well as the influence of different daytimes and measurements across continents. Moreover we a analysing the influence of IPv6 and the use of different protocols for probing in more detail.

\section*{Acknowledgment}
First of all we want to thank Sebastian Seeber for providing the needed credits for applying the measurements using RIPE Atlas. We are also grateful to Frank Tietze for assisting us with his expertise and many helpful discussions. Additional thanks for supporting us goes to RUAG Schweiz AG, Division RUAG Defence concentrating its network enabled operations activities in a business unit designated NEO Services.


\renewcommand\refname{References}
\bibliographystyle{IEEEtran}
\bibliography{IEEEabrv,literature}
\end{document}